\documentclass[twoside]{article}
\usepackage{qic,epsfig}
\usepackage{color}
\usepackage{dcolumn}
\usepackage{bm}
\usepackage{graphicx}
\usepackage{amsmath}
\usepackage{latexsym}
\usepackage{array}
\usepackage{epsfig}
\usepackage{txfonts}
\usepackage{ulem}
\usepackage{epstopdf}
\usepackage[pdftex,bookmarksnumbered,bookmarksopen,colorlinks,citecolor=blue,linkcolor=blue]{hyperref}

\newcommand{\ket}[1]{\left\vert#1\right\rangle}

\begin{document}
\setlength{\textheight}{8.0truein}    

\runninghead{CHARACTERIZATION AND PROPERTIES OF WEAKLY OPTIMAL ENTANGLEMENT WITNESSES}
            {BANG-HAI WANG, HAI-RU XU, STEVE CAMPBELL, SIMONE SEVERINI}

\normalsize\textlineskip
\thispagestyle{empty}
\setcounter{page}{1}

\copyrightheading{0}{0}{2003}{000--000}

\vspace*{0.88truein}

\alphfootnote

\fpage{1}

\centerline{\bf
CHARACTERIZATION AND PROPERTIES OF WEAKLY OPTIMAL ENTANGLEMENT WITNESSES}
\vspace*{0.37truein}
\centerline{\footnotesize
BANG-HAI WANG}
\vspace*{0.015truein}
\centerline{\footnotesize\it School of Computer Science and Technology, Guangdong University of Technology}
\baselineskip=10pt
\centerline{\footnotesize\it Guangzhou 510006,
People's Republic of China}
\centerline{\footnotesize\it Department of Computer Science, and Department of Physics and Astronomy}
\baselineskip=10pt
\centerline{\footnotesize\it University College London, Gower Street, WC1E 6BT London,
United Kingdom}
\vspace*{10pt}
\centerline{\footnotesize
HAI-RU XU}
\vspace*{0.015truein}
\centerline{\footnotesize\it School of Computer Science and Technology, Guangdong University of Technology}
\baselineskip=10pt
\centerline{\footnotesize\it Guangzhou 510006,
People's Republic of China}
\centerline{\footnotesize\it Department of Computer Science, and Department of Physics and Astronomy}
\baselineskip=10pt
\centerline{\footnotesize\it University College London, Gower Street, WC1E 6BT London,
United Kingdom}
\vspace*{10pt}
\centerline{\footnotesize
STEVE CAMPBELL}
\vspace*{0.015truein}
\centerline{\footnotesize\it Centre for Theoretical Atomic, Molecular and Optical Physics}
\baselineskip=10pt
\centerline{\footnotesize\it School of Mathematics and Physics, Queen's University
Belfast BT7 1NN, United Kingdom}
\vspace*{10pt}
\centerline{\footnotesize
SIMONE SEVERINI}
\vspace*{0.015truein}
\centerline{\footnotesize\it Department of Computer Science, and Department of Physics and Astronomy}
\baselineskip=10pt
\centerline{\footnotesize\it University College London, Gower Street, WC1E 6BT London,
United Kingdom}
\vspace*{0.225truein}
\publisher{(received date)}{(revised date)}
\vspace*{0.21truein}

\abstracts{
We present an analysis of the properties and characteristics of weakly optimal entanglement witnesses, that is witnesses whose expectation value vanishes on at least one product vector. Any weakly optimal entanglement witness can be written as the form of $W^{wopt}=\sigma-c_{\sigma}^{max} I$, where $c_{\sigma}^{max}$ is a non-negative number and $I$ is the identity matrix. We show the relation between the weakly optimal witness $W^{wopt}$ and the eigenvalues of the separable states $\sigma$. Further we give an application of weakly optimal witnesses for constructing entanglement witnesses in a larger Hilbert space by extending the result of [P. Badzi\c{a}g {\it et al}, Phys. Rev. A {\bf 88}, 010301(R) (2013)], and we examine their geometric properties.
}{}{}

\vspace*{10pt}

\vspace*{3pt}
\communicate{to be filled by the Editorial}

\vspace*{1pt}\textlineskip    
\section{Introduction}
Despite the remarkable advances in entanglement theory in the last few decades~\cite{Horodecki09,Guhne09}, we are still left with many open questions concerning it. In particular, efficient implementable means to detect, quantify, and characterize entanglement in arbitrary systems still appears a long way off. However, a significant break thorough in the practical analysis of entanglement came when it was  noticed that beyond using positive maps to characterize separability~\cite{negativity}, we can instead use Hermitian operators that then correspond to physical measurements~\cite{negativity,Jam}. Such operators were later termed entanglement witnesses (in short, witnesses)~\cite{Terhal00}. Since then a significant body of work has gone into developing this idea, both in terms of practicality, i.e. minimum experimental effort required to implement such operators, and the mathematical aspects of them~\cite{Lewenstein00,Brandao05,Gurvits02,Sperling09,Doherty04,Hou10a,Chruscinski09,Chruscinski10,Campbell,Toth}. Entanglement witnesses are now one of the most widely used tools to study entanglement both experimentally and theoretically. We refer the reader to Ref. \cite{Chruscinski14} for a recent and extensive review on entanglement witnesses.

A major advantage of these witnesses is that they do not require a complete knowledge of the state in order to determine its entanglement properties. We require only the expectation value of the state on the operator to determine if it is entangled, and therefore this can typically be done using partial knowledge of the state. A negative expectation value allows us to know with certainty that the state is entangled. However, depending on the witness, a positive expectation value does not allow us to infer anything, for instance a state that has been degraded by noise may still be entangled but a given witness is unable to detect it. This leads us to an intuitive notion (that we will define more rigorously in the proceeding sections) of optimality of entanglement witness. Put simply, an optimal witness will have a vanishing expectation value on certain separable states. Such operators were extensively studied in~\cite{Lewenstein00}. Recently, a closely related class of witnesses were proposed, weakly optimal entanglement witnesses~\cite{Badziag13}. Such operators relax the constraint on optimality by requiring the expectation value to vanish on only at least a single product vector. In this paper we explore the properties and characteristics of these operators in some detail.

\section{Preliminaries and Properties of weakly optimal entanglement witnesses}
\label{properties}
\subsection{Preliminaries}
We first formalize the necessary tools and concepts that will be examined throughout the rest of the paper. An entanglement witness is a Hermitian operator, $W=W^\dag$, such that $\text{tr}(W\sigma)\geq0$ for an arbitrary separable state $\sigma$, and there exists an entangled state $\pi$ such that $\text{tr}(W\pi)<0$. Proving that a given Hermitian operator is an entanglement witness is a difficult problem and much research has focused on consolidating the requirements to characterize the set of operators that correspond to entanglement witnesses. One such major advance in this characterization was by Lewenstein {\it et al}~\cite{Lewenstein00}, wherein they introduced the concept of optimal entanglement witnesses (OEW). Following their definitions, a witness $W$ is said to be decomposable if it can be written in the form $W=P+Q^\Gamma$ with $P, Q\geq0$, and where $\Gamma$ refers to partial transposition. If it cannot be expressed in this form it is {\it non-decomposable}. It is well known that positivity under partial transposition is a necessary and sufficient condition for separability in $2\times2$, $2\times3$, and certain $\infty\times\infty$ bipartite systems~\cite{negativity,neg}. For other dimensions it is only a sufficient condition, i.e. there exist entangled states with positive partial transpose (PPT). Non-decomposable witnesses are those which are able to detect entanglement in these positive partial transpose entangled states (PPTES). Note that for any non-negative number, $c$, if $W$ is a witness so too is $cW$ and it retains all its properties, thus we say $cW$ is as fine as $W$ \cite{Sperling09} or equivalently $c W$ is the same witness as $W$ \cite{Wang13}. For our purposes we will examine bipartite, finite dimensional entanglement witnesses.

Optimality of a witness can be expressed in a number of ways. Assuming $D_{W_1}=\{\pi\geq0$, such that $\text{tr}(W_1 \pi)<0\}$ is the set of states detected by a witness $W_1$, a second witness $W_2$ is {\it finer} if $D_{W_1}\subseteq D_{W_2}$\cite{Lewenstein00}. This then allows for the definition of an OEW: $W$ is an OEW if there exists no other finer witness. Equivalently, $W$ is an OEW if and only if $W-Q$ is no longer a witness for any $Q>0$. A witness is also optimal if it has the spanning property, that is
\begin{equation}\label{SpannedSpace}
P_W=\{|u,v\rangle\in C^m\otimes C^n: \langle u,v|W|u,v\rangle=0\},
\end{equation}
spans the whole space $C^m\otimes C^n$ \cite{Lewenstein00}. However, a remark is in order. For both decomposable and non-decomposable witnesses, there exist OEW without this spanning property~\cite{Korbicz08,Augusiak11}. Clearly determining if a given witness is an OEW is a difficult task. A necessary~\cite{Guhne09} (but not sufficient condition~\cite{Brandao05,Terhal02}) for an OEW is that there must be a separable $\sigma$ such that $\text{tr}(W\sigma)=0$. This leads us to the definition of the operators that are the focus of the remainder of this paper: A witness $W$ is called a weakly optimal entanglement witness (WOEW) if its expectation value vanishes on at least one product vector~\cite{Badziag13}. Thus the WOEW shares its definition with the $\rho$-optimal witnesses by Terhal \cite{Terhal02}.

\subsection{Properties of weakly optimal entanglement witnesses}

It should be immediately clear that OEW are special instances of WOEW. However, interestingly they still share some important properties hinting that WOEW may be easier to study, for example in the quantification of entanglement~\cite{Brandao05}. Using a witness, one can quantify the entanglement content via
\begin{equation}
\label{Fernando}
E(\rho)=\max\{0,-\min_{W\in M} \text{tr} (W\rho) \},
\end{equation}
where $M$ is the intersection of the set of entanglement witnesses with some other set $C$ such that $M$ is compact, (see~\cite{Brandao05}).
If we consider the decomposable witness
\begin{equation}
W_Q=Q_1+Q_2^\Gamma,
\label{exampleKorbicz}
\end{equation}where
\begin{equation}
Q_2=\left(\begin{array}{cccc} a & 0 & 0 & a\\
                        0 & 0 & 0 & 0\\
                        0 & 0 & 0 & 0\\
                        a & 0 & 0 & a \end{array}\right), \quad
Q_1=\left(\begin{array}{cccc} 0 & 0 & 0 & 0\\
                            0 & b & b & 0\\
                            0 & b & b & 0\\
                            0 & 0 & 0 & 0 \end{array}\right),
\end{equation}
with real positive $a$ and $b$ \cite{Wang13}. We can compute $\text{tr}(W_Q|uv\rangle\langle uv|)\!\!=\!\!0$, where $|u\rangle\!\!=\!\!\frac{\sqrt2}{2}(|0\rangle-|1\rangle)$  and $|v\rangle\!\!=\!\!\frac{\sqrt2}{2}(|0\rangle+|1\rangle)$ (see Appendix A in \cite{Wang13}). Clearly then, Eq.~\eqref{exampleKorbicz} is a WOEW, while the corresponding OEW is simply $W_Q^{opt}=Q_2^\Gamma$. We can compute the expectation values of $W_Q$ and $W_Q^{opt}$, both of which faithfully quantify the entanglement content of $\rho=\frac{1}{2}(|01\rangle+|10\rangle)(\langle 01|+\langle 10|)$. From this sense, we can also call weakly optimal witnesses as {\it local optimal witnesses} (here, ``local" does not refer to subsystems).

However, OEW and WOEW do not share all properties, in particular the following is a unique feature of WOEW.

\textit{Remark 1.} For a WOEW, $W^{wopt}$, with $\langle u,v|W^{wopt}|u,v\rangle\!=\!0$ for some $\ket{u,v}$, if $\langle u,v|P|u,v\rangle=0$ for some positive operator $P\neq0$,  ${W^{wopt'}}=W^{wopt}+P\ngeq0$, is also a WOEW. Clearly, there also exist a weakly optimal but not optimal witness $W^{wopt}$ and positive operator $Q\neq0$, for $W^{wopt}-Q$ being still a weakly optimal witness.

The operators $P\neq0$ can be split into three classes. (i) $P$ is in the space spanned by $P_{W^{wopt}}$ as Eq. (\ref{SpannedSpace}), i.e. $\langle u,v|P|u,v\rangle\!=\!0$ for a $|u,v\rangle$ with $\langle u,v|W^{wopt}|u,v\rangle\!=\!0$. In this instance either $P$ is orthogonal to the negative eigenvector space of $W^{wopt}$; or $P$ is not orthogonal to the negative eigenvector space of $W^{wopt}$ and does not cover the negative eigenvector space. (ii) $P$ is not in the space spanned by $P_{W^{wopt}}$ and does not cover the negative eigenvector space. Then there exists a product state $|u,v\rangle$ such that $\langle u,v|(W^{wopt}+P)|u,v\rangle=0$. (iii) $P$ is not in the space spanned by $P_{W^{wopt}}$ and covers the negative eigenvector space. Thus there exists at least one negative eigenvalue for $W^{wopt}+P$ and a product state $|u,v\rangle$ such that $\langle u,v|(W^{wopt}+P)|u,v\rangle=0$. Constructing these operators is more complex for cases (ii) and (iii) than for case (i) because the new space spanned by $P_{W^{wopt}+P}$ is different from the one spanned by $P_{W^{wopt}}$ in cases (ii) and (iii).

Lewenstein {\it et al}~\cite{Lewenstein00} provided the necessary and sufficient conditions for subtracting operators $Q$ from any witness $W$. Clearly, any operator $Q$ that can be subtracted from the WOEW, $W^{wopt}$, leads to $W^{wopt}-Q$ still being a WOEW. As an example, consider the witness
\begin{equation}
W=R_1+R_2^\Gamma+Q
\end{equation}
in $C^3\otimes C^3$, where
$R_1=|\phi\rangle\langle\phi|$, $R_2=|\psi\rangle\langle\psi|$, $Q=|\phi'\rangle\langle\phi'|$, $P=|\phi''\rangle\langle\phi''|$ and $|\phi\rangle=\frac{1}{\sqrt{2}}(|0\rangle|1\rangle+|0\rangle|2\rangle)$, $|\psi\rangle=\frac{1}{\sqrt{3}}(|00\rangle+|11\rangle+|22\rangle)$, $|\phi'\rangle=\frac{1}{\sqrt{2}}(|1\rangle|0\rangle+|1\rangle|2\rangle)$, $|\phi''\rangle=\frac{1}{\sqrt{2}}(|2\rangle|0\rangle+|2\rangle|1\rangle)$. We can find
\begin{equation}
W_1^{wopt}=R_1+R_2^\Gamma+P+Q=W+P,
\end{equation}and
\begin{equation}
W_2^{wopt}=R_1+R_2^\Gamma=W-Q
\end{equation}
are weakly optimal.

On the other hand, as far as Remark 1, for a WOEW (include OEW), $W^{wopt}$ with $\langle u,v|W^{wopt}|u,v\rangle\!=\!0$ for some $\ket{u,v}$, there exist some positive operator $P$ with $\langle u,v|P|u,v\rangle=0$ such that ${W^{wopt'}}=W^{wopt}+P>0$ is no longer a witness. To illustrate, consider a witness $W^{wopt}$ acting on $C^2\otimes C^3$ of the form
\begin{equation}
W^{wopt}=|\psi_+\rangle\langle \psi_+|^{T_A}+Q,
\end{equation}
where $|\psi_+\rangle=(\frac{1}{\sqrt{2}})(|00\rangle+|11\rangle)$ and $Q$ is some positive operator
acting on the Hilbert space spanned by $\{|00\rangle,|01\rangle,|10\rangle,|11\rangle\}$. It then
clearly follows that $\langle 02|W|02\rangle=0$, and so $W^{wopt}$ is WOEW. Take $P$ as
the projector onto the eigenvector of $|\psi_+\rangle\langle \psi_+|^{T_A}$ corresponding
to its negative eigenvalue. The operator $W^{wopt}+P$ satisfies $\langle 02|W^{wopt}+P|02\rangle=0$.
However, it is positive and therefore no longer a witness.

\section{Criteria for weakly optimal entanglement witnesses}
\label{criteria}
To characterize the requirements for a witness to be a WOEW we make use of the results from~\cite{Wang13,Wang11} where it is shown that any (possibly unnormalized) witness $W$ can be written as
\begin{equation}
W=\sigma-c_\sigma I,
\label{ew-form0}
\end{equation}
where $\sigma$ is a separable density matrix, $\lambda_{0\sigma}<c_\sigma\leq c_\sigma^{max}$ is a real number related to $\sigma$, $\lambda_{0\sigma}$ is the minimum eigenvalue of $\sigma$ and
\begin{equation}
c_\sigma^{max}=\inf_{\parallel|\mu_A\rangle\parallel=1,
\parallel|\mu_B\rangle\parallel=1}\langle\mu_A\mu_B|\sigma|\mu_A\mu_B\rangle
\end{equation}
is the maximum number in $c_\sigma$ which makes $W=\sigma-c_\sigma I$ a witness with $|\mu_A\mu_B\rangle$ any unit product vector.

This implies that the characterization of any entanglement witness $W$ is equivalent to characterizing the (separable) density matrix $\sigma$ and determining $c_\sigma^{max}$. From here we are now ready to determine the criteria for a witness to be a WOEW.

\textit{Remark 2.} A witness $W=\sigma-c_{\sigma} I$ is weakly optimal if and only if $c_{\sigma}=c_\sigma^{max}$.

Therefore, we have a general procedure to construct weakly optimal witnesses from other witnesses. (i) We can compute $c_\sigma^{max}$ such that $W^{wopt}=\sigma-c_\sigma^{max}I$ for any witness $W=\sigma-c_\sigma I$. (ii) We can add $Q\geq0$ to $W^{opt}=\sigma-c_\sigma^{max}I$ such that $c_{\sigma}^{max}=c_{\sigma+Q}^{max}$ and $W^{wopt}=\sigma+Q-c_{\sigma+Q}^{max}I$ for an optimal witness $W^{opt}$.

Determining $c_\sigma^{max}$ is in general a difficult problem, despite an algorithm for its calculation within a finite number of steps exists~\cite{jamiolkowski}, which was recently reformulated in terms of Lagrangian multipliers~\cite{Badziag13}. Considering the fineness associated with witnesses, we know the expectation values for WOEW coincidentally vanish on only a subset of the states detected by the OEW.
In the following we can restrict ourselves to witnesses of the form Eq.~\eqref{ew-form0}, for $c_{\sigma}=c_{\sigma}^{max}$. We will now show that $c_{\sigma}^{max}$ is closely related to the eigenvalues of the separable matrix $\sigma$ for decomposable and non-decomposable witnesses.

\vskip0.2cm
\noindent
\textit{Theorem 1.} If $W^{dopt}=\sigma-c_\sigma^{max} I$ is a decomposable optimal witness, $c_\sigma^{max}=\lambda_{0\sigma^\Gamma}$, where $\lambda_{0\sigma^\Gamma}$ is the minimum eigenvalue of $\sigma^\Gamma$.
\vskip0.2cm

In order to prove this we require the following lemma~ \cite{Lewenstein00,Korbicz08}.

\textit{Lemma 1.} Any decomposable optimal entanglement witness $W^{dopt}$ can be written as $W^{dopt}=Q^\Gamma$, where $Q\geq0$, and there exits no $P$ in the range of $Q$ such that $P^\Gamma\geq0$.

{\it Proof:}
It then follows from Lemma 1 the partial transpose of the decomposable OEW can be expressed,  $(W^{dopt})^\Gamma=\sigma^\Gamma-c_\sigma^{max} I\geq0$ with $c_\sigma^{max} \leq \lambda_{0\sigma^\Gamma}$. However since $W^{dopt}=\sigma-c_\sigma^{max} I$ is an OEW, we have $\lambda_{0\sigma} < c_\sigma^{max}$. This implies $\lambda_{0\sigma}<\lambda_{0\sigma^\Gamma}$, and $W=\sigma-\lambda_{0\sigma^\Gamma}I<0$. Since $(\sigma-\lambda_{0\sigma^\Gamma}I)^\Gamma=\sigma^\Gamma-\lambda_{0\sigma^\Gamma}I\geq0$, we have that $W=\sigma-\lambda_{0\sigma^\Gamma}I$ is also a witness and $\lambda_{0\sigma^\Gamma}\leq c_\sigma^{max}$. Therefore $c_\sigma^{max}=\lambda_{0\sigma^\Gamma}$.  \hfill$\square$

Let us demonstrate the above theorem by an example. Consider the normalized decomposable OEW
\begin{eqnarray}
W^{wopt}&=&
    \left(
      \begin{array}{cccc}

                 0.5 &0       & 0      &0  \\

                 0   &0     &0.5 & 0 \\

                 0   &0.5 &0     &0 \\

                 0   &0       &0       &0.5
      \end{array}
    \right)
    \label{SpecialCase}\nonumber\\
    &=&\sigma_1-0.5I\nonumber\\
    &=&\sigma_2-0.6I,
\end{eqnarray}
where $\sigma_1$ and $\sigma_2$ are the following (unnormalized) separable matrices
\begin{equation}
\sigma_1=\left(\begin{array}{cccc} 1.0 & 0 & 0 & 0\\
                        0 & 0.5 & 0.5 & 0\\
                        0 & 0.5 & 0.5 & 0\\
                        0 & 0 & 0 & 1.0 \end{array}\right), \quad
\sigma_2=\left(\begin{array}{cccc} 1.1 & 0 & 0 & 0\\
                            0 & 0.6 & 0.5 & 0\\
                            0 & 0.5 & 0.6 & 0\\
                            0 & 0 & 0 & 1.1 \end{array}\right).
\end{equation}
We can compute the minimum eigenvalue of ${\sigma_1}^\Gamma$ finding $\lambda_{0{\sigma_1}^\Gamma}=0.5$, and the minimum eigenvalue of ${\sigma_2}^\Gamma$, $\lambda_{0{\sigma_2}^\Gamma}=0.6$. From the method outlined in Ref. \cite{Wang13} we can find $c_{\sigma_1}^{max}=0.5$, and $c_{\sigma_2}^{max}=0.6$. This demonstrates the above theorem as $c_{\sigma_1}^{max}=\lambda_{0{\sigma_1}^\Gamma}$ and $c_{\sigma_2}^{max}=\lambda_{0{\sigma_2}^\Gamma}$.

Following Ref. \cite{Wang13,Wang11}, we can also obtain the dual witness form of Eq.~\eqref{ew-form0},
\begin{equation}\label{ew-form00}
W=c_\sigma I-\sigma,
\end{equation}
where $\sigma$ is a separable, normalized density matrix, with $c_\sigma$ a real number related to $\sigma$ and satisfying $c_\sigma^{min}\leq c_\sigma <\lambda_{M\sigma}$. Here $\lambda_{M\sigma}$ is the maximum eigenvalue of $\sigma$ and
\begin{equation}
\label{ew-form1}
c_\sigma^{min}=\sup_{\parallel|\mu_A\rangle\parallel=1,
\parallel|\mu_B\rangle\parallel=1}\langle\mu_A\mu_B|\sigma|\mu_A\mu_B\rangle
\end{equation}
is the minimum number for $c_\sigma$ which makes $W=c_\sigma I-\sigma$ a witness, when $|\mu_A\mu_B\rangle$ is any unit product vector. This then leads to the following corollary of theorem 1.

\textit{Corollary 1.} If $W^{dopt}=c_\sigma^{min} I-\sigma$ is a decomposable optimal entanglement witness, $c_\sigma^{min}=\lambda_{M\sigma^\Gamma}$, where $\lambda_{M\sigma^\Gamma}$ is the maximum eigenvalue of $\sigma^\Gamma$.

Clearly, either there exists $(W^{dwopt})^\Gamma\ngeq0$ or $(W^{dwopt})^\Gamma\geq0$ for some decomposable WOEW, ${W^{dwopt}}$. We then have the following corollary.

\textit{Corollary 2.} For a decomposable weakly optimal entanglement witness $W^{dwopt}=\sigma-c_\sigma^{max} I$, if $(W^{dwopt})^\Gamma\geq0$, $c_\sigma^{max}=\lambda_{0\sigma^\Gamma}$; if $(W^{dwopt})^\Gamma\ngeq0$, $W=\sigma-tI$ and $W^\Gamma=\sigma^\Gamma-tI$ are witnesses for $\max\{\lambda_{0\sigma},\lambda_{0\sigma^\Gamma}\}<t<c_\sigma^{max}$.

In the case of non-decomposable entanglement witnesses we have the following.

\textit{Remark 3.} A non-decomposable witness, $W$, is a WOEW if and only if $W^\Gamma$ is weakly optimal.

This can be seen by considering that for any non-decomposable witness $W\ngeq0$, its partial transpose, $W^\Gamma\ngeq0$, otherwise $W$ can not detect PPTESs. Let us assume that $W$ is not weakly optimal. Then $\langle u|\langle v|W|u\rangle|v\rangle>0$ for any $|u\rangle|v\rangle$. However, as $W$ is non-decomposable then there exists at least one $|u\rangle|v\rangle$ such that the expectation value of its partial transpose, $W^\Gamma$, $\langle u|\langle v|W^\Gamma|u\rangle|v\rangle=0$, i.e. $W^\Gamma$  is an non-decomposable WOEW. However, this means $\langle u|\langle v^*|W|u\rangle|v^*\rangle=0$, which contradicts our initial assumption and therefore $W$ is an indecomposable WOEW, thus satisfing the `if' part of the statement. If there exist a product state $|u\rangle|v\rangle$ such that $\langle u|\langle v|W|u\rangle|v\rangle=0$, there must exist a product state $|u\rangle|v^*\rangle$ such that $\langle u|\langle v^*|W^\Gamma|u\rangle|v^*\rangle=0$. That is, if $W$ is an indecomposable weakly optimal witness, $W^\Gamma$ is also weakly optimal. Satisfing the `only if' part of the statement.

\textit{Corollary 3.} If $W^{ndwopt}=\sigma-c_\sigma^{max} I$ is a non-decomposable weakly optimal entanglement witness, $c_\sigma^{max}=c_{\sigma^\Gamma}^{max}$.

\section{The application of weakly optimal witnesses: Construction of Entanglement Witnesses in a Larger Product Hilbert space}

Recently, Badzi{\c{a}}g {\it et al} considered any separable state $\sigma\in B(\mathcal{H}_A\otimes\mathcal{H}_B)$, and constructed an associated WOEW $W_\sigma$ on a larger product Hilbert Space $\mathcal{H}'\otimes\mathcal{H}'$ with $\text{dim}(\mathcal{H}')\leq (d_Ad_B)^4$ ($\text{dim}(\mathcal{H}_A\otimes\mathcal{H}_B)=(d_Ad_B)^2)$ \cite{Badziag13}. They concluded the following main result.

{\it Lemma 2}~\cite{Badziag13}. A bipartite state $\sigma$ is separable if and only if its corresponding witness
\begin{equation}\label{Witness-form}
W_\sigma=Y+C_YP^{asym},
\end{equation}
with $C_Y>\parallel Y\parallel_\infty$ is a WOEW, where
$Y=P^{sym}AP^{sym}$, $P^{sym}=[(1/2)(I\otimes I+V\otimes V)]$, $P^{asym}=[(1/2)(I\otimes I-V\otimes V)]$,
\begin{equation}\label{Degree-four-form}
A=\alpha B\otimes P_{cl}+\beta I\otimes(V-P_{cl})+\gamma I\otimes(P_{cl}-I/N),
\end{equation} with
\begin{equation}
\mathcal{B}(\sigma)=\text{min}_{u\in H'\otimes H'}\langle u,u|A|u,u\rangle>0
\end{equation}
the biconcurrence function, and $B$ the associated biconcurrence matrix \cite{Badziag02}, $P_{cl}=\sum_i|i\rangle|i\rangle\langle i|\langle i|$, $V=\sum_{ij}|i\rangle |j\rangle\langle j|\langle i|$ is the swap operator, $I$ is the identity matrix, and $\alpha,\beta,\gamma>0$. Moreover, if $C\geq2\parallel Y\parallel_\infty$, $\langle u,v|W_\sigma|u,v\rangle\geq\mathcal{B}(\sigma)$.

Interestingly we can use any state $\rho\in B(\mathcal{H}_A\otimes\mathcal{H}_B)$ to construct a witness in larger size than $W_\sigma$ by substituting $B'=\rho\otimes\rho\otimes\rho\otimes\rho$ for $B$ in Eq. (\ref{Degree-four-form}). We can construct the witnesses acting on $C^{N^4}\otimes C^{N^4}$ with $N=(d_Ad_B)^2$.

{\it Corollary 4.}  For any state $\rho\in B(\mathcal{H}_A\otimes\mathcal{H}_B)$,
\begin{equation}\label{W-from-state}
W_\rho=Y'+C_{Y'}P'^{asym}
\end{equation}
is a witness acting on
a product Hilbert space $(\mathcal{H}_{AB}\otimes\mathcal{H}_{AB}\otimes\mathcal{H}_{AB}\otimes\mathcal{H}_{AB})\otimes(\mathcal{H}_{AB}\otimes\mathcal{H}_{AB}\otimes\mathcal{H}_{AB}\otimes\mathcal{H}_{AB})$, where $Y'=P'^{sym}A'P'^{sym}$ and
\begin{eqnarray}\label{A}
A'&=&\alpha B'\otimes P'_{cl}+\beta I\otimes(V'-P'_{cl})+\gamma I\otimes(P'_{cl}-I'/(N^2))\end{eqnarray},
where $B'=\rho\otimes\rho\otimes\rho\otimes\rho$ and $P'_{cl}, V', I'$ are acting on $\mathcal{H}_{AB}\otimes\mathcal{H}_{AB}\otimes\mathcal{H}_{AB}\otimes\mathcal{H}_{AB}$.

Following precisely the arguments of Badzi{\c{a}}g {\it et al}~\cite{Badziag13}, we find a practical witness in the same dimension as $W_\sigma$ by substituting $Y'=\frac{1}{2}(W\otimes W\otimes W\otimes W+(W\otimes W\otimes W\otimes W)V\otimes V)$ for $Y$, where $W\in B(\mathcal{H}_A\otimes\mathcal{H}_B)$ is any witness. To prove this result, we need the following lemma.

{\it Lemma 3}~\cite{Badziag13}. Let $X$ be a Hermitian operator acting on a product Hilbert space $H\otimes H$ such that
 \begin{equation}\label{X}
 X=P^{sym}XP^{sym}
 \end{equation}
 and $\langle u,u|X|u,u\rangle\geq0$ for any $|u\rangle\in H$. Let $X_C=X+C_XP^{asym}$, where C is a real constant. The following implications are true:(i) if $C\geq\parallel X\parallel_\infty$, $\langle u,v|X_C|u,v\rangle\geq0$ for any pair of vectors $|u\rangle,|v\rangle\in H$, and if $C\geq2\parallel X\parallel_\infty$,there exists $|g\rangle\in H$ such that
\begin{eqnarray*}
  \langle u,v|X_C|u,v\rangle &\geq& \langle g,g|X_C|g,g\rangle \\
                             &\geq& \text{inf}_{|u\rangle\in H}\langle u,u|X|u,u\rangle (=:X).
\end{eqnarray*}

{\it Remark 4.}  For any witness $W\in B(\mathcal{H}_A\otimes\mathcal{H}_B)$, we can construct a witness
\begin{equation}\label{W-W}
W_W=Y''_C=Y''+C_{Y''}P^{asym}
\end{equation}
acting on
a product Hilbert space $(\mathcal{H}_{AB}\otimes\mathcal{H}_{AB})\otimes(\mathcal{H}_{AB}\otimes\mathcal{H}_{AB})$, where $Y''=P^{sym}Y'P^{sym}$ and $Y'=\frac{1}{2}(W\otimes W\otimes W\otimes W+(W\otimes W\otimes W\otimes W)V\otimes V)$.

{\it Proof:} For any $|u\rangle\in\mathcal{H}_{AB}\otimes\mathcal{H}_{AB}$,
\begin{eqnarray}
&&\langle u,u|Y'|u,u\rangle\nonumber\\
&=&\frac{1}{2}\langle u,u|W\otimes W\otimes W\otimes W|u,u\rangle+\frac{1}{2}\langle u,u|(W\otimes W\otimes W\otimes W)V\otimes V|u,u\rangle\nonumber\\
&=&\frac{1}{2}(\langle u|W\otimes W|u\rangle)^2+\frac{1}{2}(\langle u|(W\otimes W)V|u\rangle)^2 \label{Expectation}\nonumber\\
&\geq&0.
\end{eqnarray}

Following the steps of~\cite{Badziag13}, we circumvent the fact that $Y'$ is not a witness (due to the fact it has negative expectations values for some product vectors $|u,v\rangle$ when $|u\rangle\neq|v\rangle$) by adding a suitably weighted projection onto the antisymmetric subspace $P^{asym}=[(1/2)(I\otimes I-V\otimes V)]$. We then replace $Y'$ by this new operator $Y''=P^{sym}Y'P^{sym}$, and this substitution does not affect the expectation values:

By $V(W\otimes W)=W\otimes W$ and $V^2=I$,
\begin{eqnarray*}
&&P^{sym}Y'P^{sym}\\
 &=&P^{sym}(\frac{1}{2})(W\otimes W\otimes W\otimes W+(W\otimes W\otimes W\otimes W)V\otimes V)P^{sym}\\
 &=& (\frac{1}{4})(\frac{1}{2})(I\otimes I+V\otimes V)(W\otimes W\otimes W\otimes W+(W\otimes W\otimes W\otimes W)V\otimes V)(I\otimes I+V\otimes V)\\
 &=& (\frac{1}{4})(\frac{1}{2})(W\otimes W\otimes W\otimes W+(W\otimes W\otimes W\otimes W)V\otimes V\nonumber\\
 &&+(V\otimes V)(W\otimes W\otimes W\otimes W+(W\otimes W\otimes W\otimes W)V\otimes V))\times(I\otimes I+V\otimes V)\\
 &=&(\frac{1}{4})(\frac{1}{2})(2W\otimes W\otimes W\otimes W+2(W\otimes W\otimes W\otimes W)V\otimes V)(I\otimes I+V\otimes V)\\
 &=&(\frac{1}{4})(\frac{1}{2})(2W\otimes W\otimes W\otimes W+2(W\otimes W\otimes W\otimes W)V\otimes V\nonumber\\
 &&+(2W\otimes W\otimes W\otimes W+2(W\otimes W\otimes W\otimes W)V\otimes V)(V\otimes V))\\
 &=&(\frac{1}{4})(\frac{1}{2})(2W\otimes W\otimes W\otimes W+2(W\otimes W\otimes W\otimes W)V\otimes V\nonumber\\
 &&+(2W\otimes W\otimes W\otimes W(V\otimes V)+2(W\otimes W\otimes W\otimes W)V^2\otimes V^2))\\
 &=&(\frac{1}{4})4(\frac{1}{2})(W\otimes W\otimes W\otimes W+(W\otimes W\otimes W\otimes W)V\otimes V)\\
 &=& Y'.
\end{eqnarray*}

Lemma 3 then tells us the weight which $P^{asym}$ has to be added to $Y''$ to make it an entanglement witness since it has at least one negative eigenvalue.
\hfill$\square$

For example consider the witness
\begin{eqnarray}
W=\left(
     \begin{array}{cccc}

                 \frac{1}{8} &0       & 0      &-\frac{1}{4}  \\

                 0   &\frac{3}{8}     &0 & 0 \\

                 0   &0     &\frac{3}{8}    &0 \\

                 -\frac{1}{4}   &0       &0       &\frac{1}{8}
      \end{array}
    \right).
\end{eqnarray}
$W_W=\frac{1}{2}(W\otimes W\otimes W\otimes W+(W\otimes W\otimes W\otimes W)V\otimes V)-\frac{162}{4096}(I\otimes I-V\otimes V)$ is a witness acting on $C^{16}\otimes C^{16}$.

Both $W_\sigma$ in Eq. (\ref{Witness-form}) and $W_W$ in Eq. (\ref{W-W}) are bipartite witnesses of the same dimension. One is constructed from a separable state, $W_{\sigma}$, and  the other is constructed from a witness. While $W_\sigma$ must be a WOEW, the nature of $W_W$ depends on the properties of the original witness $W$, if $W$ is a WOEW then $W_W$ must also be weakly optimal.


\section{Relevant Examples}
Here we make use of the above statements to two examples. First, we apply them to the Werner state~\cite{Wang11,Werner89}
\begin{equation}
\label{werner}
\pi_p=p|\psi\rangle\langle\psi|+(1-p)I/4,
\end{equation}
where $|\psi\rangle=\frac{1}{\sqrt{2}}(|00\rangle+|11\rangle)$ and $0\leq p\leq1$. It is well known that Eq.~\eqref{werner} is entangled if $p>\frac{1}{3}$. We can construct a witness
\begin{equation}
\label{isotropic-witness}
W=\sigma_q-\frac{1+q}{4}I
\end{equation}
to detect Eq.~\eqref{werner}, where
\begin{equation}
\sigma_q=q|\psi\rangle\langle\psi|+(1-q)I/4=
    \left(
      \begin{array}{cccc}

                 \frac{1+q}{4}  &0 & 0 &\frac{q}{2}  \\

                 0 & \frac{1-q}{4} &0 & 0 \\

                 0  &0 & \frac{1-q}{4} &0 \\

                 \frac{q}{2}&0 &0 &\frac{1+q}{4}
      \end{array}
    \right)
\end{equation}
for $-\frac{1}{3}<q<0$. We can determine tr$(W\pi_p)=\frac{(3p-1)q}{4}<0$ for $1>p>\frac{1}{3}$ and $-\frac{1}{3}<q<0$. Since $\frac{1+3q}{4}<\frac{1+2q}{4}<\frac{1+q}{4}$, $W_1=\sigma_q-\frac{1+2q}{4}I$ can detect $\pi_p$ for $p>\frac{2}{3}$, where $\frac{1+3q}{4}$ is the minimum eigenvalue of $\sigma_q$. We see that Eq.~\eqref{isotropic-witness} is finer than $W_1=\sigma_q-\frac{1+2q}{4}I$ and it is an OEW.

We can also construct a WOEW
\begin{equation}\label{isotropic-witness1}
W'=\sigma'_q-\frac{1+q}{4}I
\end{equation}
to detect Eq.~\eqref{werner}, where
\begin{equation}
\sigma'_q=
    \left(
      \begin{array}{cccc}

                 \frac{1+q}{4}  &0 & -\frac{q}{2} &\frac{q}{2}  \\

                 0 & \frac{1-q}{4} &0 & 0 \\

                 0  &0 & \frac{1-q}{4} &0 \\

                 \frac{q}{2}&-\frac{q}{2} &0 &\frac{1+q}{4}
      \end{array}
    \right)
\end{equation}
for $-\frac{1}{3}<q<0$.

{Similarly for higher dimensions, consider the family of witnesses in $\mathbb{C}^3 \otimes \mathbb{C}^3$
defined by \cite{Chruscinski09,Cho-Kye}
\begin{equation}\label{}
  W[x,y,z]\, =\,  \left( \begin{array}{ccc|ccc|ccc}
    x & \cdot & \cdot & \cdot & -1 & \cdot & \cdot & \cdot & -1 \\
    \cdot& y & \cdot & \cdot & \cdot& \cdot & \cdot & \cdot & \cdot  \\
    \cdot& \cdot & z & \cdot & \cdot & \cdot & \cdot & \cdot &\cdot   \\ \hline
    \cdot & \cdot & \cdot & z & \cdot & \cdot & \cdot & \cdot & \cdot \\
    -1 & \cdot & \cdot & \cdot & x & \cdot & \cdot & \cdot & -1  \\
    \cdot& \cdot & \cdot & \cdot & \cdot & y & \cdot & \cdot & \cdot  \\ \hline
    \cdot & \cdot & \cdot & \cdot& \cdot & \cdot & y & \cdot & \cdot \\
    \cdot& \cdot & \cdot & \cdot & \cdot& \cdot & \cdot & z & \cdot  \\
    -1 & \cdot& \cdot & \cdot & -1 & \cdot& \cdot & \cdot & x
     \end{array} \right)\ ,
\end{equation}
where $x,y,z\geq 0$ and $\cdot$ denotes $0$. Necessary and sufficient conditions for
$W[x,y,z]$ to be a witness are
\begin{equation}\label{EW-condition}
\begin{array}{lll}
 (a)& 0 \leq x < 2,&\, \\
 (b)& x+y+z \geq 2,&\,\\
 (c)& \quad\mbox{if}\, x \leq 1, \quad\mbox{then}\,&  yz\geq (1-x)^2
 \end{array}\biggl\rbrace
\end{equation}
A family $W[x,y,z]$ generalizes Choi
indecomposable witnesses corresponding to $x=y=1$ and $z=0$,
\begin{equation}\label{}
  W[1,1,0]= \left( \begin{array}{ccc|ccc|ccc}
    1 & \cdot & \cdot & \cdot & -1 & \cdot & \cdot & \cdot & -1 \\
    \cdot& 1 & \cdot & \cdot & \cdot& \cdot & \cdot & \cdot & \cdot  \\
    \cdot& \cdot & 0 & \cdot & \cdot & \cdot & \cdot & \cdot &\cdot   \\ \hline
    \cdot & \cdot & \cdot & 0 & \cdot & \cdot & \cdot & \cdot & \cdot \\
    -1 & \cdot & \cdot & \cdot & 1 & \cdot & \cdot & \cdot & -1  \\
    \cdot& \cdot & \cdot & \cdot & \cdot & 1 & \cdot & \cdot & \cdot  \\ \hline
    \cdot & \cdot & \cdot & \cdot& \cdot & \cdot & 1 & \cdot & \cdot \\
    \cdot& \cdot & \cdot & \cdot & \cdot& \cdot & \cdot & 0 & \cdot  \\
    -1 & \cdot& \cdot & \cdot & -1 & \cdot& \cdot & \cdot & 1
     \end{array} \right).
\end{equation}
We can calculate eigenvalues of $W[1,1,0]$, $\lambda_0=-1$, $\lambda_1=\lambda_2=\lambda_3=0$, $\lambda_4=\lambda_5=\lambda_6=1$ and $\lambda_7=\lambda_8=2$ and eigenvectors $|v_0\rangle=\frac{1}{\sqrt{3}}(|00\rangle+|11\rangle+|22\rangle)$, $|v_1\rangle=|02\rangle$, $|v_2\rangle=|10\rangle$, $|v_3\rangle=|21\rangle$, $|v_4\rangle=|01\rangle$, $|v_5\rangle=|12\rangle$, $|v_6\rangle=|20\rangle$, $|v_7\rangle=\frac{1}{\sqrt2}(|11\rangle-|00\rangle$, and $|v_8\rangle=\frac{1}{\sqrt2}(|22\rangle-|00\rangle)$.

Since $W[1,1,0]$ is optimal \cite{Ha2012}, we can construct many weakly optimal witnesses. Let $Q_0=\frac{1}{2}|v_0\rangle\langle v_0|+|v_1\rangle\langle v_1|$, $Q_1=|v_2\rangle\langle v_2|+|v_7\rangle\langle v_7|$ then $W_1=W[1,1,0]+Q_0$ and $W_2=W[1,1,0]+Q_1$ are weakly optimal. We can write $W[1,1,0]$ as
\begin{equation}\label{}
  W[1,1,0]=\sigma-2I,
\end{equation}
where
\begin{equation}\label{}
  \sigma= \left( \begin{array}{ccc|ccc|ccc}
    3 & \cdot & \cdot & \cdot & -1 & \cdot & \cdot & \cdot & -1 \\
    \cdot& 3 & \cdot & \cdot & \cdot& \cdot & \cdot & \cdot & \cdot  \\
    \cdot& \cdot & 2 & \cdot & \cdot & \cdot & \cdot & \cdot &\cdot   \\ \hline
    \cdot & \cdot & \cdot & 2 & \cdot & \cdot & \cdot & \cdot & \cdot \\
    -1 & \cdot & \cdot & \cdot & 3 & \cdot & \cdot & \cdot & -1  \\
    \cdot& \cdot & \cdot & \cdot & \cdot & 3 & \cdot & \cdot & \cdot  \\ \hline
    \cdot & \cdot & \cdot & \cdot& \cdot & \cdot & 3 & \cdot & \cdot \\
    \cdot& \cdot & \cdot & \cdot & \cdot& \cdot & \cdot & 2 & \cdot  \\
    -1 & \cdot& \cdot & \cdot & -1 & \cdot& \cdot & \cdot & 3
     \end{array} \right).
\end{equation}
Since $\lambda_{0\sigma}=1$, all $W=\sigma-cI$ are witnesses for $1<c\leq2$. Letting $c=1.5$, $W=\sigma-1.5I$ is the same witness as $W[1.5,1.5,0.5]$, which satisfies Eq. (\ref{EW-condition}).} Since $W[1,1,0]$ is non-decomposable,

\begin{equation}\label{}
  W[1,1,0]^\Gamma=\sigma^\Gamma-2I
\end{equation}
is also an optimal witness \cite{Lewenstein00}, where
\begin{equation}\label{}
  \sigma^\Gamma= \left( \begin{array}{ccc|ccc|ccc}
    3 & \cdot & \cdot & \cdot & \cdot & \cdot & \cdot & \cdot & \cdot \\
    \cdot& 3 & \cdot & -1 & \cdot& \cdot & \cdot & \cdot & \cdot  \\
    \cdot& \cdot & 2 & \cdot & \cdot & \cdot & -1 & \cdot &\cdot   \\ \hline
    \cdot & -1 & \cdot & 2 & \cdot & \cdot & \cdot & \cdot & \cdot \\
    \cdot & \cdot & \cdot & \cdot & 3 & \cdot & \cdot & \cdot & \cdot  \\
    \cdot& \cdot & \cdot & \cdot & \cdot & 3 & \cdot & -1 & \cdot  \\ \hline
    \cdot & \cdot & -1 & \cdot& \cdot & \cdot & 3 & \cdot & \cdot \\
    \cdot& \cdot & \cdot & \cdot & \cdot& -1 & \cdot & 2 & \cdot  \\
    \cdot & \cdot& \cdot & \cdot & \cdot & \cdot& \cdot & \cdot & 3
     \end{array} \right).
\end{equation}

We can easily compute $c_\sigma^{max}=c_{\sigma^\Gamma}^{max}=2$, all $W=\sigma^\Gamma-cI$ are witnesses for $\frac{5-\sqrt5}{2}<c\leq2$. Interestingly, $W=\sigma^\Gamma-cI$ is non-decomposable for $1<c\leq2$, but it is decomposable for $\frac{5-\sqrt5}{2}<c\leq1$.

\section{The geometric properties of weakly optimal entanglement witnesses}
\label{geometric}
Following from the considerations of Bertlmann {\it et al}~\cite{Bertlmann05} in this section we consider some of the geometrical aspects of WOEW. In the finite dimensional Hilbert space $\mathcal{H}_{AB}=\mathcal{C}^{d_{A}}\otimes\mathcal{C}^{d_{B}}$, if $d_A=d_B=d_{AB}$ the total dimension of $\mathcal{H}_{AB}$ will be $d_{AB}^2$ (as considered in~\cite{Bertlmann02} for example). Clearly any witness $W$ must correspond to an observable given by a Hermitian matrix, with states expressed as density matrices, $\varrho$, which are themselves combinations of vectors in $\mathcal{H}_{AB}$. We can regard these quantities as elements of a real Hilbert space $\mathcal{H}_r=\mathcal{R}^{d_{AB}^4}$ with scalar product given by
\begin{equation}
\langle\varrho|W\rangle=\text{tr}\left( \varrho W \right)
\end{equation}
and corresponding norm
\begin{equation}
\|W\|_2=(\text{tr}W^2)^{\frac{1}{2}}.
\end{equation}

\begin{figure}[t]
\centerline{\epsfig{file=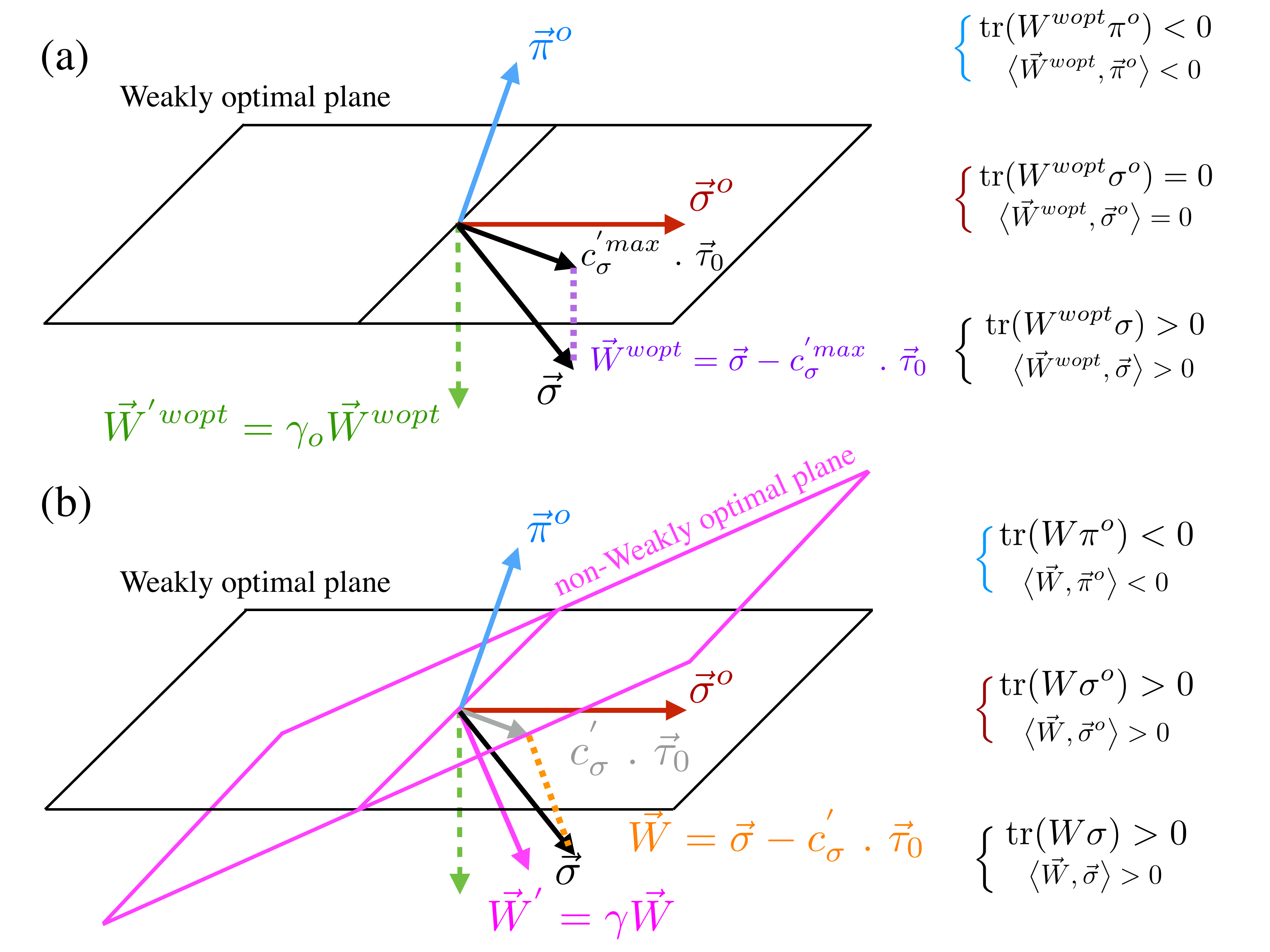,width=.55\columnwidth}}
\fcaption{(Color online) Geometric illustration of (a) Weakly optimal entanglement witness $W^{wopt}$ and the weakly optimal plane $\vec{W}^{wopt}$; (b) The non-weakly optimal and weakly optimal entanglement witnesses $W^{nwopt}$, $W^{wopt}$ and the non-weakly-optimal and weakly-optimal planes, $\vec{W}$ and $\vec{W}^{wopt}$, respectively. See main text for discussions.}
\label{fig1}
\end{figure}

By the Hahn-Banach theorem, we can view a witness, $W$, as a hyperplane with dimension $d_{AB}-1$, which has
the entangled state $\pi$, which is detected by $W$, on one side while all separable state lie in the
hyperplane or on the other side~\cite{Pittenger02}. We can rewrite Eq.~\eqref{ew-form0} as
\begin{equation}
W=\sigma-c'_\sigma\tau_0,
\label{ew-density0-sub}
\end{equation}
where $\tau_0=\frac{1}{d_{AB}}I$ is the maximally mixed state. This means any witness can be expressed as the difference between a separable state, $\sigma$, and the product of a non-negative real number with the maximally mixed state. In Euclidean space we can represent the WOEW and non-WOEW with simple planes, as shown in Fig.~\ref{fig1} (a) and (b). Hyperlanes separate ``left-hand" entangled states from ``right-hand" and ``inside" separable states. In panel (a) the witness $W^{wopt}=\sigma-c'^{max}_\sigma\tau_0$ is the same WOEW as $W'^{wopt}=\gamma_o W^{wopt}$, where $\sigma$ is a separable state, $\tau_0$ is the maximally mixed state, and $c'^{max}_\sigma$ and $\gamma_o$ are non-negative numbers. The vector $\vec{W}^{wopt}=\vec{\sigma}-c'^{max}_\sigma\vec{\tau_0}$ denotes the same plane as the parallel $\vec{W}^{'wopt}$. We also see $\pi^o$ is entangled such that tr$(W^{wopt}\pi^o)<0$, while $\sigma^{o}$, which is inside the weakly optimal hyperlane, is separable such that tr$(W^{wopt}\sigma^{o})=0$. In panel (b) we also compare with the hyperplane associated with a non-WOEW.

\begin{figure}[t]
\centerline{\epsfig{file=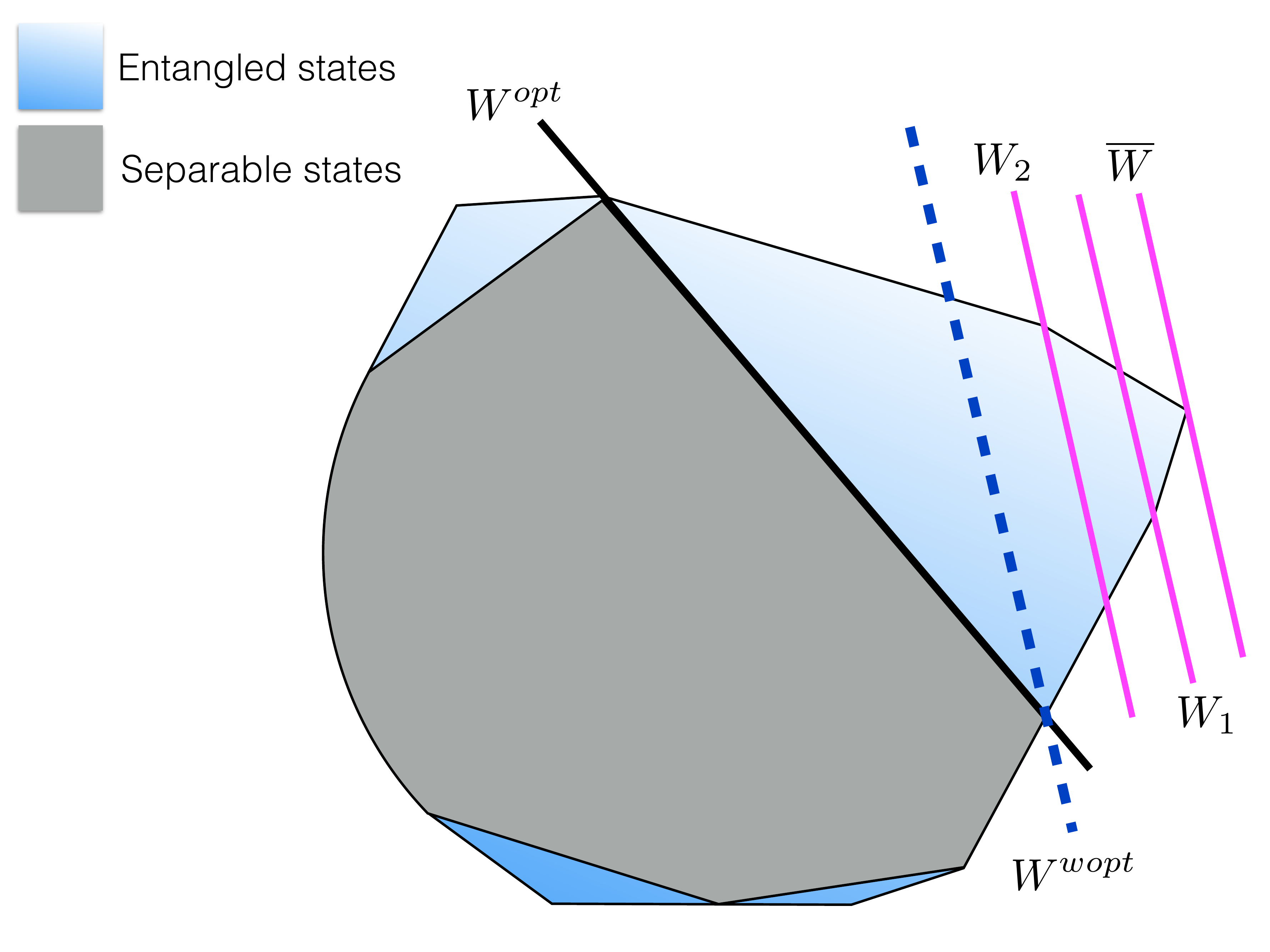,width=.55\columnwidth}}
\fcaption{(Color online) Schematic representation of the relations between OEW, WOEW, and finer witnesses in the sets of entangled and separable states.}
\label{fig2}
\end{figure}

The relation between separable states, entangled states and different witnesses in the form $W=\sigma-c_\sigma I$ is shown in Fig.~\ref{fig2}. Each witness corresponds to a hyperplane in Hermitian operator space similar to those shown in Fig.~\ref{fig1}. $W^{opt}=\sigma'-c_\sigma'^{max} I$ is an OEW, while $W^{wopt}=\sigma-c_\sigma^{max} I$ is a WOEW. We also show other witnesses illustrating the {\it fineness} property, $W_2=\sigma-c_{2\sigma} I$ is finer than $W_1=\sigma-c_{1\sigma} I$. The ``boundary" of witnesses, $\bar{W}=\sigma-\lambda_{0\sigma} I$ is not an entanglement witness~\cite{Wang11}, where $\lambda_{0\sigma}$ is the minimum eigenvalue of separable state $\sigma$ satisfying Eq. (\ref{ew-form0}). We can see that generally, OEWs can be viewed as tangent hyperplanes to the set of separable states with a WOEW acting as a supporting hyperplane.

\section{Conclusions}
\label{conclusions}
We have presented an analysis of weakly optimal entanglement witnesses (WOEW), that is witnesses whose expectation value vanishes on at least one product vector. We have shown how these operators can be easier to obtain and study than their optimal counterparts and while still providing a useful tool to study entanglement. Interestingly, WOEW are closely related to the eigenvalues of complementary separable matrices also showing a method to construct these entanglement witnesses in a larger Hilbert space by any witness or any quantum state and we have explored their geometrical properties.

\nonumsection{Acknowledgements}
\noindent
We are grateful to Fernando Brand\~{a}o and Dong-Yang Long for very helpful discussions and suggestions. We want to thank the referee for extremely careful reading and many comments that have led to substantial improvements. We carried on this work while Wang was an academic visitor and Xu an undergraduate visiting student of the Department of Computer Science and the Department of Physics \& Astronomy at University College London. Wang and Xu would like to thank this institution for the kind hospitality. This work is supported by the National Natural Science Foundation of China under Grant No. 61272013 and the National Natural Science Foundation of Guangdong province of China under Grant No. s2012040007302.

\nonumsection{References}

\end{document}